\documentstyle[sprocl,amssymb,epsf]{article}

 \newcommand{\bm}[1]{\mbox{\boldmath $#1$}}

 \newcommand{\zr}[1]{\mbox{\hspace*{#1em}}}

 \newcommand{\Id}{\mbox{1\zr{-0.62}{\small 1}}}

 \newcommand{\QQ}{\mbox{{\sf Q}\zr{-0.5}\rule[0.1ex]{0.04em}{1.4ex}\zr{0.53}}}

 \newcommand{\TT}{{\Bbb T}}
 \newcommand{\RR}{{\Bbb R}}

\begin{document}

\title{SYMMETRY STRUCTURE OF THE ELSER-SLOANE QUASICRYSTAL  }

\author{\sc Michael Baake {}\footnote{Heisenberg-fellow
        \quad\quad ${}^b$ supported by the Swiss BBW (HCM programme)}}

\address{Institut f\"ur Theoretische Physik, Universit\"at T\"ubingen, \\
Auf der Morgenstelle 14, D-72076 T\"ubingen, Germany}

\author{\sc Franz G\"ahler \negthinspace${}^b$} 
%\footnote{supported by the Swiss BBW
%        (HCM programme)}
%}

\address{Centre de Physique Th\'eorique, Ecole Polytechnique,\\
F-91128 Palaiseau, France}


\maketitle\abstracts{
The 4D quasicrystal of Elser and Sloane, obtained from the root lattice 
$E_8$ by the cut-and-project method, 
can be parametrized by the points of an 8D torus. 
This allows for an explicit analysis of its point 
and inflation symmetry structure. }

\section{Introduction}

It is well known that the local isomorphism class (LI-class) of a
crystallographic pattern $\cal P$ in $\RR^n$ (i.e.\ a pattern whose periods
span $\RR^n$) consists of $\cal P$ and its translates. 
Since the translate ${\cal P} + \ell$ equals $\cal P$ for any $\ell$
in the lattice $\Lambda$ of periods of $\cal P$, the LI-class is in
one-to-one correspondence with points of a fundamental domain of
$\Lambda$ which (on identifying opposite facets) is an $n$-dimensional
torus, $\TT^n$. This can now be used to find all patterns in LI($\cal P$)
with special symmetries (relative to the origin)
which is a standard procedure in crystallography.

{}For repetitive, but non-crystallographic $\cal P$, however, the LI-class
has a much richer structure and contains uncountably many ($2^{\aleph^{}_0}$)
translation classes. It is thus much more difficult to classify the complete
symmetry structure of such classes, and no general answer is known.
If the pattern happens to be quasi-crystallographic (in the sense that it
stems from a standard projection scheme), the key to parametrizing its
LI-class is to use the fundamental domain of the 
embedding lattice. This is the so-called {\em torus parametrization}
that has been introduced recently~\cite{BHP} and then applied
to some of the most frequently used quasicrystallographic 
tilings~\cite{BHP,HRB} in two and three dimensions.

It is the aim of this contribution to extend this set of examples to the
Elser-Sloane quasicrystal in four dimensions.~\cite{ES}
It is constructed by the projection method from the root lattice $E_8$
and has the Coxeter group $H_4$ of order 14400 as its point symmetry group,
together with an inflation/deflation symmetry with scaling factor
$\tau = (1+\sqrt{5})/2$. It is of interest due to its role in the
hierarchy of quasicrystals with $\tau$ inflation.
%Below, we shall briefly describe the structure of inflation invariant
%tilings in this LI-class, as well as those with special point symmetries.

\section{Setup and application to inversion symmetry}

In what follows, we shall use the notation and the main results of
Ref.~1 without further reference. We standardize our
parametrization such that the origin of the embedding space is at 
a lattice point of $E_8$, and the window (having point symmetry $H_4$)
is symmetric around the origin of internal space. 
With this convention, the tiling obtained by cut-and-project
with the physical space cutting through the origin has parameter
$\bm{t}=\bm{0}$. 

To each symmetry operation on the LI-class now corresponds a linear 
(or, more generally, affine) operator
on the torus, and tilings with special properties can be found as
fixed points of that operator on the torus. Also, the number of solutions
can be counted: an equation of the form
\begin{equation}
      A \bm{x} + \bm{t} \; = \; \bm{x} \quad {\rm mod} \; \TT^8
\end{equation}
has precisely  $|\det(A-\Id\,)|$ different solutions, provided 1 is not
an eigenvalue of $A$ (for details, also on the singular case,
see the appendix of Ref.~1).

Let us illustrate this with the simplest case, that of inversion symmetry.
In view of our standardization, the corresponding operator in 8-space is
defined by the isometry $\bm{x} \mapsto -\bm{x}$. Restricting this to the
torus, we are thus asking for the number of solutions of
$\bm{t} = -\bm{t}$ mod $\TT^8$, which is $|\det(-2\cdot\Id\,)| = 2^8 = 256$.
They are the 2-division points of the torus (solving $2 \bm{t}=\bm{0}$ on it).
We shall see a little later how these points (and hence the tilings
parametrized by them) are distributed over other properties.

%Let us now mention one subtlety suppressed so far:
There is one subtlety which has been suppressed so far:
the parametrization is one-to-one only for regular
members of the LI-class (those where the cut space never hits the boundary
of the window), while it is multiple-to-one for the remaining singular
tilings, which are then grouped into classes.~\cite{BHP}

\section{Inflation structure}

One of the most interesting properties of non-periodic tilings is
the existence of inflation/deflation symmetries, and in the present
case there is one with inflation multiplier $\tau$ (the golden ratio).
To describe this, we start with the corresponding
situation for the Fibonacci chain, discussed
in detail in Ref.~1. It is shown there that one has $a^{}_n$ Fibonacci
chains invariant under $n$-fold inflation, where
\begin{equation}
   a^{}_n \; = \; a^{}_{n-1} + a^{}_{n-2} + 1 - (-1)^n
\end{equation}
with initial values $a^{}_1 = a^{}_2 = 1$. This follows from the
determinant argument discussed above and can directly be calculated
from the eigenvalues $\tau$ and $-1/\tau$ of the corresponding operator
on the torus ($\TT^2$ in this case).

This result can directly be used for our present task.
The number of points on the 8-torus that are invariant under $n$-fold
inflation (denoted by $I^n$) is given by the fourth power of the 
number $a^{}_n$, because we get the same eigenvalues as before,
but now with multiplicity 4.
Next, let us denote by $b_n^{(4)}$ the number of fixed points under $I^n$
which are not fixed by $I^m$ for any $m<n$. It can recursively be calculated
by
\begin{equation}
       b_n^{(4)} \; = \; a_n^4 \; - \sum_{m|n,m<n} b_m^{(4)} \, .
\end{equation}
These numbers correspond to points that can be grouped into $n$-cycles, so
$c_n^{(4)} = b_n^{(4)}/n$ must be an integer. Table 1 lists these numbers
for $1\leq n\leq 8$.

\begin{table}[ht] 
\centerline{
\begin{tabular}{|c|rrrrrrrr|}   \hline
$n$         &  1&  2&  3&  4&   5&   6&  7&  8 \\
\hline
$a_n^4$     &  1&  1&  256&  625&  14641&  65536&  707281&  4100625 \\
$b_n^{(4)}$ &  1&  0&  255&  624&  14640&  65280&  707280&  4100000 \\
$c_n^{(4)}$ &  1&  0&   85&  156&   2928&  10880&  101040&   512500 \\
\hline
\end{tabular}  }
\caption{Inflation orbit counts of the Elser-Sloane tiling. They also 
apply to any other 4D cut-and-project patterns with $\tau$-inflation.} 
\label{icotab}
\end{table} 

To summarize these findings, one can attach a dynamical $\zeta$-function 
to the inflation operator $I$ on the torus. It reads
\begin{equation}
     Z_4(x) \, = \, 
            \frac{(1-4x-x^2)^4 (1+4x-x^2)^4 
                  (1-x-x^2)^{28} (1+x-x^2)^{28}}
                 {(1-7x+x^2) (1-3x+x^2)^{12} (1+3x+x^2)^{16}
                  (1+x)^{32} (1-x)^{38}} \, . \!
\end{equation}
This serves as generating function for the counts $a^4_n$ via
\begin{equation}
    \log(Z_4(x)) \; = \; \sum_{n=1}^{\infty} a^4_n \cdot {x^n\over n}
\end{equation}
and -- through an Euler product decomposition -- also for the counts $c_n^{(4)}$:
\begin{equation}
{1\over Z_4(x)} \; = \; \prod_{n=1}^\infty(1-x^n)^{c_n^{(4)}} \; = \;
  (1-x)^1(1-x^2)^0(1-x^3)^{85}(1-x^4)^{156}\ldots
\end{equation}
This $\zeta$-function has the functional equation $Z_4(x)=Z_4(1/x)$
and conforms to the ``Riemann hypothesis'' that its zeros $\alpha$
satisfy $N_2[\alpha] = -1$ where $N_2$, the norm of the quadratic field 
$\QQ(\tau)$, is defined by $N_2[r + s \tau] = r^2 + r s - s^2$.

Let us point out here that the 256 inversion symmetric tilings of section 
2 coincide with the 256 tilings invariant under $I^3$. This is so because
$\tau^3 = 2 \tau + 1$ and the corresponding equations on the torus define
the same set of solutions.

\section{Space group and point group structure}

\begin{figure}[b!]
\vskip -0.4cm
\centerline{\epsfysize=90mm  \epsffile{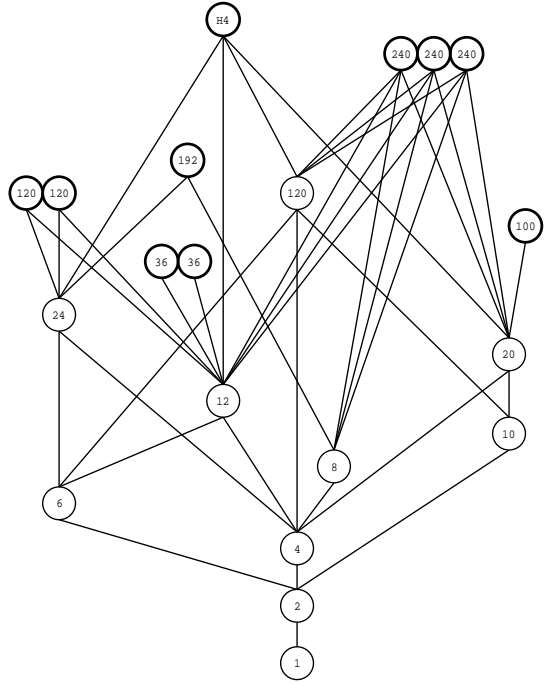} }
\vskip -0.1cm
\caption{Classification of torus points according to their
  space group symmetry. Each vertex corresponds to a 
  Wyckoff position, i.e., a class of points whose stabilizers are 
  conjugate subgroups of the space group of the 8D periodic
  structure. The order of the stabilizer is inscribed into each vertex
  (except for $H_4$). Incidence relations of Wyckoff positions are 
  indicated by connecting lines (those with smaller stabilizers
  contain others with bigger stabilizers). A Wyckoff position
  consists of a space group orbit of some (rational) affine subspace, 
  wrapped on the torus. Vertices for Wyckoff positions of dimension 
  zero (special points) are drawn with thick lines. They are
  discretely embedded on the torus. Each space group orbit of special 
  points corresponds to one circle. Circles which touch each other 
  represent orbits of special points with the same point group symmetry. }
\end{figure}

So far, except inversion symmetry, we have only discussed inflation symmetry, 
which does not depend on point group symmetry.
But the window of our quasicrystal is a special polytope~\cite{ES} with $H_4$ 
symmetry, and so the generalized point group of the LI-class is $H_4$,
and its space group is the semi-direct product of $H_4$ with the 
limit translation module of the tiling. The action of this space group can 
easily be lifted to an affine action on the torus $\TT^8$. Since the 
space group is symmorphic, its action on the torus, where one computes
modulo lattice or module translations, in fact agrees with that
of the point group, so that we need not distinguish between space 
group and point group action.

Let us now see how single members of the LI-class reflect this point 
group structure, seen through their torus parameters. To this end, 
we have computed the set of Wyckoff positions and their incidence 
relations for our quasicrystal (see Fig.~1 and its caption). 
This was done with a program called CrystGap.~\cite{EGN,FG} 
On the torus, Wyckoff positions consist of unions of subtori of
various dimensions, whose points have conjugate stabilizers.
Fig.~1 shows the hierarchy of these subtori (which is closely
related to the classification of lower-dimensional 
examples~\cite{BHP}), together with their incidence
relations. The latter also imply subgroup relations 
for their stabilizers, which make up an only small but essential part of 
the (very complex) subgroup lattice of $H_4$. The point 
groups contained in Fig.~1 as stabilizer of some point on the 
torus (and thus of some tiling) make up only a small fraction of the 
set of all subgroups of $H_4$. 

Some points with non-trivial point group symmetry also have 
non-trivial inflation symmetry. Here, we shall discuss the inflation 
symmetry only for the special points, which are discrete on the torus. 
They all have a unique inflation symmetry. 
The only point with $H_4$ symmetry coincides with that 
of $\tau$ inflation ($I$) symmetry. Points in the 
three orbits of length 60 (stabilizer size: 240), as well as points
in the orbit of length 75 (192), have $I^3$ symmetry. In fact, the 
points in these orbits make up all 255 such points. The three orbits 
of length 60 are permuted by $I$, whereas the 75-orbit is reshuffled 
internally by $I$. The two orbits of length 120 (120) and the orbit 
of length 144 (100) are $I^4$ symmetric. The pair of 120-orbits is 
permuted by $I^2$. Finally, the pair of orbits of length 400 (36) 
is permuted by $I^4$; all of its points are invariant under $I^8$. 

Conversely to the case of $I^3$ symmetry, the special
points do not make up all points with $I^8$ or even $I^4$
symmetry. The lacking points are to be found on the other tori,
which have positive dimension. These contain, in fact, a dense set 
of points (all rational points) with some inflation symmetry, 
although of varying degree. For this reason we have confined 
ourselves to special points.

\end{document}